\documentstyle[aas2pp4]{article}
\newcommand{\beq}{\begin{equation}}
\newcommand{\eeq}{\end{equation}}
\newcommand{\etal}{et~al.}

\begin{document}

\title{Interferometric Astrometry of \\ the Low-mass Binary Gl 791.2
(= HU Del)\\ Using {\it Hubble Space Telescope} Fine Guidance Sensor
3: \\ Parallax and Component Masses
\footnote{Based on observations made with the NASA/ESA Hubble Space
Telescope, obtained at the Space Telescope Science Institute, which is
operated by the Association of Universities for Research in Astronomy,
Inc., under NASA contract NAS5-26555} }

\author{G.\ Fritz Benedict\altaffilmark{1}, Barbara E.
McArthur\altaffilmark{1}, Otto G. Franz\altaffilmark{2}, Lawrence\
H. Wasserman\altaffilmark{2}, and Todd J. Henry\altaffilmark{3}}

\altaffiltext{1}{McDonald Observatory, University of Texas, Austin, TX 78712}
\altaffiltext{2}{Lowell Observatory, 1400 West Mars Hill Rd., Flagstaff, AZ 86001}
\altaffiltext{3}{Department of Physics and Astronomy, Johns Hopkins University, Baltimore, MD 21218}

\begin{abstract}

With fourteen epochs of fringe tracking data spanning 1.7y from Fine
Guidance Sensor 3 we have obtained a parallax ($\pi_{abs}=113.1\pm0.3$
mas) and perturbation orbit for Gl 791.2A. Contemporaneous fringe
scanning observations yield only three clear detections of the
secondary on both interferometer axes. They provide a mean component magnitude difference, $\Delta V =
3.27\pm0.10$. The period (P = 1.4731 yr) from the perturbation orbit
and the semi-major axis ({\it a} = 0.963 $\pm$ 0.007 AU) from the
measured component separations with our parallax provide a total system mass ${\cal M}_A
+ {\cal M}_B = 0.412 \pm 0.009 {\cal M}_{\sun}$. Component masses are
${\cal M}_A=0.286 \pm 0.006 M_{\sun}$ and ${\cal M}_B = 0.126\pm0.003
{\cal M}_{\sun}$.  Gl 791.2A and B are placed in a sparsely populated
region of the lower main sequence mass-luminosity relation where they
help {\it define} the relation because the masses have been determined
to high accuracy, with errors of only 2\%.

\end{abstract}


\section{Introduction}

The dependence of intrinsic brightness upon mass, the mass-luminosity
relation (MLR), is fundamental to our understanding of stellar
astronomy.  The MLR is relevant to studies of the evolution of
individual stars, and is required to convert a luminosity function
into a mass function for determinations of the mass content of the
Galaxy.  The MLR can also be folded into efforts to determine the ages
of stellar clusters.  In addition, at masses less than 0.1 ${\cal
M}_{\sun}$ the MLR is critical for brown dwarf studies, because mass
remains an important determinant in culling brown dwarfs from stars.
Despite its broad utility, the MLR remains poorly determined for M
dwarfs, by far the dominant population of the Galaxy in both numbers
($>$70\%) and stellar mass contribution ($>$40\%; Henry 1998).  To
improve the low-mass MLR, we are observing nearby red dwarf multiple
systems in order to determine masses accurate to 5\% or better.

Gl 791.2 (HU Del = LHS 3556, RA = 20 29 48.0 DEC = +09
   41 23, epoch and equinox 2000) is an example of an M dwarf binary with at least one
component in the 20-20-20 Sample, which was defined by Henry et al. (1999) to
include systems having d $\le$ 20 pc, ${\cal M} \le 0.20{\cal
M}_{\sun}$, and orbital period P $\le$
20 years. Gl 791.2 was discovered to be binary by \cite{Har71}, who
determined a semi-major axis for the photocentric orbit of 0\farcs029
with a period of 1.5 yr, and estimated the companion mass to be
0.07-0.11 ${\cal M}_{\sun}$.  The perturbation was confirmed by
\cite{Her78}, who found nearly identical values for the semi-major
axis, period, and mass estimate, and suggested that ``Such a pair may
be resolvable with a space telescope. One reliable observation of
separation and magnitude would yield good masses for both members.''

Using infrared speckle techniques \cite{McC86}, attempted to resolve
the pair but was unsuccessful.  Continued attempts by one of us
(Henry) using infrared speckle imaging detected the companion
($\Delta$H (1.6$\mu$m) $\leq$ 1.3) but did not result in clear
resolution because of the close separation (maximum $\sim$0\farcs15).
The observation did indicate that the magnitude difference was not
extreme (presumably even at optical wavelengths), so that the system
might be resolved with Fine Guidance Sensor 3 (FGS 3) on {\it HST}.
Most importantly, FGS 3 offered resolution capability to 20 mas, about
10 times better than the infrared speckle observations.  With the
suspicion that the secondary is an important object with a mass near
$0.1{\cal M}_{\sun}$ in a small but resolvable orbit, we obtained 14
usable fringe tracking observation sets with {\it HST} to obtain the precise parallax and
component masses reported in this paper.

Table~\ref{tbl-1} provides aliases and physical parameters for the Gl
791.2 system. The radius is predicted from the model of \cite{Bur93},
and is comparable with the radii of the red dwarf components of CM Draconis, 0.25$R_{\sun}$ and 0.23$R_{\sun}$, determined by \cite{Met96}, which have
masses of 0.23${\cal M}_{\sun}$ and 0.21${\cal M}_{\sun}$.  Finally, \cite{Del98} find
Gl 791.2 to be a rapid rotator and from space velocities assign it to the young disk population. 

\section{Data Reduction and Calibration Procedures}

Our observations were obtained with FGS 3, a two-axis, white-light
interferometer aboard {\it HST}.  This instrument has two operating modes; fringe tracking (POS) and fringe scanning (TRANS). We time-tag our data with a modified
Julian Date, $MJD = JD - 2400000.5$, and abbreviate millisecond of
arc, mas, throughout. We obtained a total of fifteen observation sets. Unfortunately the POS observations at mJD 50534.04 suffered from over 50 mas of drift. Satisfactory correction was unobtainable, so we discarded these data. Table~\ref{tbl-2} presents a log of the
successful POS observations used in this investigation. Note that each observation
set contains either two or three observations of the primary science
target, Gl 791.2, because a single long visit has been split into
shorter, independent measurements. Table~\ref{tbl-3} presents a log of those
TRANS observations that yielded total component separations, position angles, and magnitude differences. These TRANS data are discussed in Section 2.3.

\cite{Bra91}, provide an overview of the FGS 3 instrument. Our goal, 1 mas per-observation precision small-field astrometry, has been achieved, but not without significant challenges. These included a mechanically noisy on-orbit environment, the self-calibration of FGS 3, and significant temporal changes in our instrument. 
\cite{Ben99}, and \cite{Ben98}, review our solutions to these problems, 
including a denser set of drift check stars for each science observation, fine-tuning exposure times, overlapping field observations and analyses for calibration, and a continuing series of trend-monitoring observations. \cite{Mca99}, and \cite{Ben99}, describe data acquisition and reduction strategies for derivation
of parallax and proper motion. 

The present investigation poses the
added complexity of reflex motion of the target due to a faint
companion. Hence our GaussFit (\cite{Jef88}) model was modified to
allow for the simultaneous estimation of the astrometric parameters
($\xi$, $\eta$), parallax ($\pi$), and proper motion ($\mu$$_x$,
$\mu$$_y$), as well as the seven standard orbital parameters (see
Table 6). We first place all the reference star measurements into a
common reference frame by determining scale, orientation, and offset
parameters for each observation set. These are 'plate' constants. We then apply
these to the Gl 791.2A measurements to obtain the parallax, proper
motion, and orbital elements.

\subsection {The Gl 791.2 Astrometric Reference Frame}  

\label{AstRefs} 
\begin{figure}
\epsscale{1.0}
\plotone{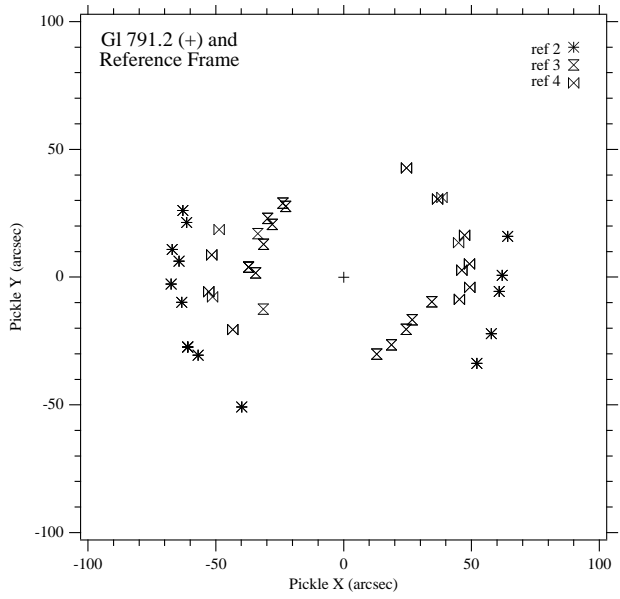}
\caption{Gl 791.2 and reference
frame observations in FGS 3 pickle coordinates. The symbol shape
identifies each star listed in Table 3.}
\label{fig-1}

\end{figure}

Table~\ref{tbl-4} lists the three stars in the Gl 
791.2 reference frame.  Figure \ref{fig-1} shows the distribution in
FGS 3 instrumental coordinates of the 14 sets of reference star
measurements used to establish the reference frame. The circular
pattern is impressed by the requirement that {\it HST} roll to keep
its solar panels fully illuminated throughout the year.

From these data we determine the scale, rotation, and offset ``plate
constants" relative to an arbitrarily adopted constraint epoch (the so-called ``master plate") for
each observation set. With a sufficient number
of reference stars we normally obtain six plate constants per
observation set, allowing the scale in X to differ from the scale in
Y. We also determine the parallax and proper motion of each reference star. The Gl 791.2 reference frame contains only three stars. Hence, we
constrained the scales along X and Y to equality and the two axes to orthogonality. We also constrained the proper motions and parallaxes to have values of zero. The model becomes

\beq
\xi = aX + bY + c 
\eeq 
\beq 
\eta = -bX +aY +f 
\eeq 

\noindent where a, b, c and f are the plate constants, relating each observation set to the 'master plate'.

The orientation with respect to the sky of the observation set chosen
to be the constraint epoch is determined with star positions from the
USNO-A2.0 catalog (\cite{Mon98}) with uncertainties in the field
orientation $\pm 0\fdg03$.

\begin{figure}
\epsscale{0.6}
\plotone{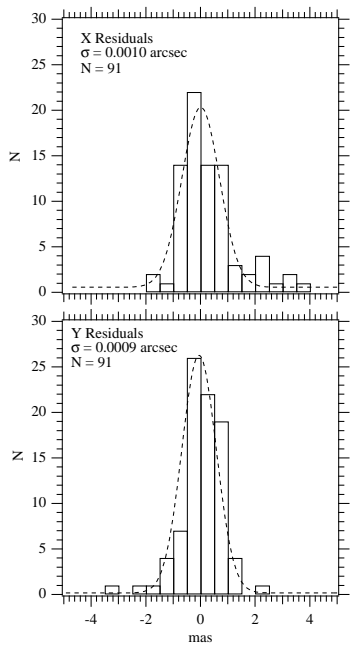}
\caption{Histograms of x and y residuals obtained from modeling the Gl
791.2 reference frame with equations 1 and 2. Distributions are fit
with gaussians.} \label{fig-2}
\end{figure}

The final reference frame model produces plate constants from the 91
reference star observations contained in the 14 observation sets. (Each set contains one to three observations of each reference star.) It also produces residuals for each reference star. From the
histograms of the X and Y residuals (Figure~\ref{fig-2}) we conclude
that we have obtained plate constants that map each observation set to
the others at the  1 mas level. To determine if there might be
unmodeled, but eventually correctable, systematic effects, we plotted
the Gl 791.2 reference frame X and Y residuals against a number of
spacecraft, instrumental, and astronomical parameters. These included
X and Y position within the FGS field of regard (the so-called pickle), radial distance from the pickle
center, reference star V magnitude and $(B-V)$ color, and epoch of
observation.  No trends were seen above the 1.0 mas level.  The resulting catalog of positions for the three reference stars is given in Table~\ref{tbl-4}, where $\xi$ and $\eta$ are parallel to RA and Dec, respectively.

\subsection {Modeling the Parallax, Proper Motion, and Orbital Motion 
of Gl 791.2A}

Once we have determined plate constants from applying equations 1 and
2 to the reference frames, we apply the transformations to the Gl
791.2A fringe tracking measurements (FGS POS mode) and solve for
parameters describing relative parallax, proper motion, and orbital
motion, modifying the equations appropriately.

\beq
\xi = aX + bY + c - P_x*\pi - \mu_x*t - ORBIT_x
\eeq 
\beq 
\eta = -bX +aY +f - P_y*\pi - \mu_y*t - ORBIT_y
\eeq 

\noindent where a, b, c and f are the plate constants obtained from the reference frame, $P_x$ and $P_y$
are the parallax factors, and t is the time of observation. We obtain the parallax factors, $P_x$ and
$P_y$, from a JPL Earth orbit predictor (DE200, \cite{Sta90}). ORBIT is a function of the traditional orbital elements, listed in Table 6.

From the photometry in Tables 1 and 4 we see that the colors of the
reference stars and the target are all red. Therefore, we apply no
corrections for lateral color (\cite{Ben99}).  Reference frame colors
were obtained from the USNO-A2.0 catalog (Monet 1998). Calibration
from the catalog b-r colors to B-V was obtained from the Barnard's
Star and lateral color calibration field photometry in Benedict et
al., 1999. We obtained a zero point correction from the Leggett (1992)
photometry of Gl 791.2.

Given the uncorrected HST point spread
function (PSF) for the FGSs (they are not in the COSTAR path), we must discuss
 photocenter corrections (e.g., \cite{vdK67}). The
FGS transforms the images of the two components into two fringes. As
long as the components are more widely separated than the resolution
of {\it HST} at $\lambda = 580$ nm, about 40 mas, the presence of a
companion has no effect on the centroid obtained from the fringe
zero-crossing. However, what matters is the separation along 
each interferometer axis, not the total separation
on the sky. For any arbitrary orientation components A and B can have 
separations along the FGS axes from 0.0 mas to the actual
full separation. For example on 1996d084 
(observation set 6 in Table 2) the total separation
along the X axis is 151 mas, but only 26 mas along Y.

Simulations were done to characterize the effect the secondary has
 on the position derived for the primary for separations
less than 40 mas.  First, we obtained
a position derived from a fringe scan of a single star of appropriate color.
Then, a companion with $\Delta$V = 3.25 was placed at separations of
$\pm5, \pm 10, \pm 20, \pm 35, \pm 50$ and $\pm 100$ mas along each axis. As can be seen in \cite{Fra98}, (fig. 1), the interferometer response functions
are not symmetric about the zero crossing. Hence the simulation required
that the companion be placed on either side of the primary.
Fitting each synthetic binary TRANS scan as a single star will quantify the effect of the undetected companion on the measured position of the primary.
In no case did the position of the primary move from the single star location by more than 0.6 mas, less than the error in the measurement of its position.

Every small-field astrometric technique requires a correction from
relative to absolute parallax because the reference frame stars have
an intrinsic parallax.  We adopt the methodology discussed in the Yale
Parallax Catalog (\cite{WvA95}, Section 3.2, Fig. 2, hereafter
YPC95). From YPC95, Fig. 2, the Gl 791.2 galactic latitude, $b =
-16\fdg8$, and average magnitude for the reference frame,
$<V_{ref}>=13.9$, we obtain a correction to absolute of 1.0 $\pm$ 0.2
mas. Applying this correction results in the absolute parallax of
112.9 $\pm$ 0.3 mas listed in Table~\ref{tbl-5}.  This compares
favorably to the YPC95 parallax, 113.8 $\pm$ 1.9 mas, and represents a
six-fold reduction in the error.  With V = 13.06, Gl 791.2 is too
faint to have been observed by {\it HIPPARCOS}.

The final absolute parallax formal uncertainty includes the estimated
error in the correction to absolute, added in quadrature to the
relative parallax error. Our confidence in the correction to absolute
($\pm0.2$ mas) comes from comparisons carried out by \cite{Har99}, and
\cite{Ben99}. They compared corrections to absolute derived from
spectroscopy of reference frame stars with the correction obtained
from YPC95. Differences averaged around 0.2 mas.

\subsection {The Orbit from Fringe Tracking and Scanning Observations}

If the secondary in a binary system can be detected directly, both the
magnitude difference between the components (needed to measure the
luminosity for the MLR) and their total separation (needed to measure
the mass for the MLR) can be determined. We detect only component A of
Gl 791.2 in the POS measurements, but Gl 791.2B was
detected at three epochs using FGS 3 TRANS measurements. These data are acquired at the mid-point of each
observation set (orbit) listed in the observation log (Table~\ref{tbl-2}). Details of the fringe scanning analysis procedure are presented in Franz et al
(1998).

The observed fringe is treated as a linear superposition of two single
star fringes. Two single star fringe templates are fit to the observed
binary system fringe pattern, minimizing the residuals in a
least-squares sense, with $\Delta F583W$ (F583W is the name of the
filter used) and component separation as free parameters on each axis.
The X and Y axes are fit independently because there are, in fact, two
interferometers within each FGS. Fringe scanning observations were
obtained during each of the 14 observation sets listed in Table 2. The
template used for deconvolution of both sources was from a single star
with $(B-V)$ = 1.9, similar to both components of Gl 791.2.

We obtained only three pairs of detections of the secondary in 30 attempts
(counting each axis as a separate attempt), indicating that component
B is a challenging target for FGS 3. All of these detections were
near maximum elongation, and the companion was clearly seen on both axes.  Measuring the full separation and position angle requires
resolution along both the X and Y axes at each epoch.  
At two other
epochs near maximum elongation (sets 6 and 7, Table~\ref{tbl-2}) a separation was measured, but required that we constrain the
magnitude difference on one axis. This was due to the projected separation along that axis being less than 25 mas. Because constraining $\Delta
F583W$ does not yield an independent detection in these cases, we have
removed these epochs from further consideration. 

The successful two axis detections are detailed in Table~\ref{tbl-3} and plotted with the
component A perturbation orbit in Figure~\ref{fig-4}. We note that the X
and Y positions used to derive the fringe scan separations (in
Table~\ref{tbl-3}, $\rho$ for the total separation) have been
corrected for the instrumental effect discussed in Franz et al (1998,
\S4.1). 

\begin{figure}
\epsscale{1.0}
\plotone{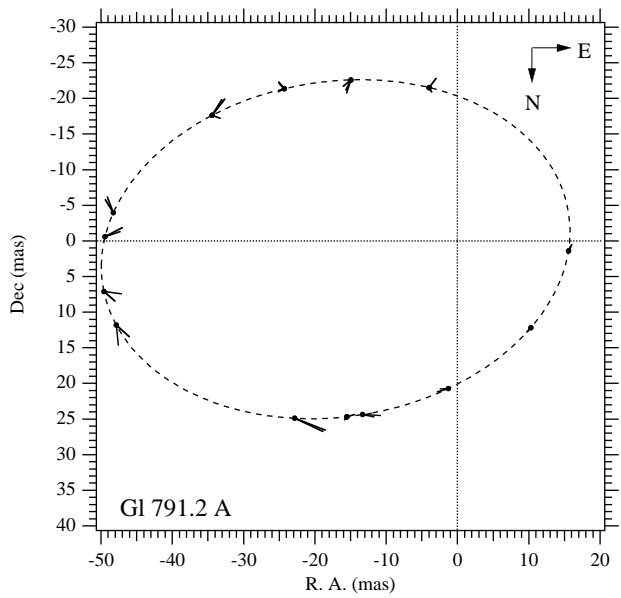}
\caption{Perturbation orbit for Gl
791.2A. Elements are found in Table~\ref{tbl-6}. Dots are predicted
positions at each epoch of observation. Residual vectors are plotted
for each observation, two or three at each epoch; see
Table~\ref{tbl-2}).} \label{fig-3}

\end{figure}

\begin{figure}
\epsscale{1.0}
\plotone{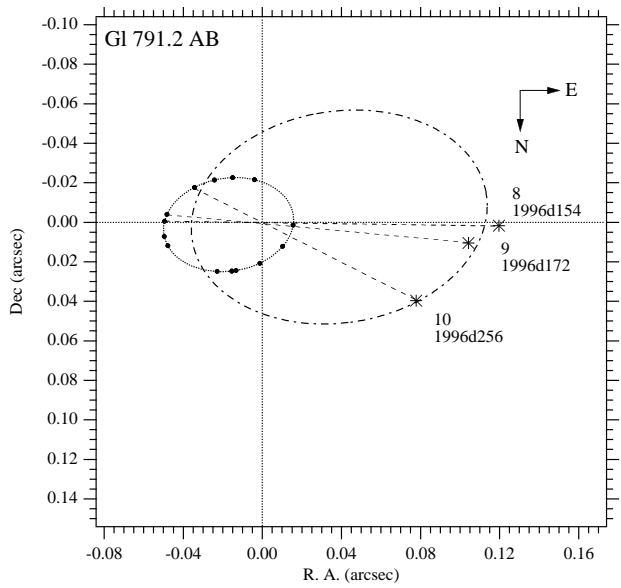}
\caption{Gl 791.2A (dots, POS measurements) and
component B (asterisks, TRANS detections). Component B
positions are labeled with their corresponding observation set numbers
(Table~\ref{tbl-3}). All observations, POS and TRANS, were used to derive the Table 6 orbital elements.} \label{fig-4}

\end{figure}

Our orbital parameters for components A and B are presented in
Table~\ref{tbl-6}. All parameters describing this orbit are derived from all the component A POS measurements (Table 2) and the three independent TRANS observations (Table 3) in a simultaneous solution. Primed (') parameters refer to the perturbation orbit. In Figure~\ref{fig-3} we show the
residual vectors for each of the two or three POS mode observations at each
epoch.  Listed in Table~\ref{tbl-2}, the residuals to the perturbation orbit range
from 0--5 mas with an average absolute value residual, $<|res|> = 1.0$
mas. The three TRANS mode separations and position angles are plotted with the 
derived component A and B orbits in Figure~\ref{fig-4}. We note that the three vectors connecting the two components intersect very near the center of mass.

\subsection{Component Magnitude Difference}
For the three epochs listed in Table 3 we were able to obtain a
reliable measure of $\Delta m$ (more precisely, $\Delta F583W$) for component
separations along both the X and Y axis. As discussed in \cite{Hen99},
the $\Delta F583W$ value along the X axis is usually smaller than that
along the Y axis, presumably because the X axis transfer function is
degraded relative to Y.  We have therefore derived a correction using 63 FGS 3 observations of 12 systems (\cite{Hen99}). This correction,
amounting to +0.08 mag, is far less than the obvious scatter in
our measurements. Nonetheless, applying the
correction to only the $\Delta F583W_X$, we obtain the $\Delta m_X^c$ values shown in Table 3. Averaging both axes we obtain a final $\Delta
F583W$ = 3.32 $\pm$ 0.10, for which we have adopted the error as
described in \cite{Hen99}.  Finally, a slight conversion is required
from the F583W filter to the V band, as described in the same paper,
and we have a final value of $\Delta V$ = 3.27 $\pm$ 0.10.

\section{Gl 791.2 Total and Component Masses}

The orbit (Table~\ref{tbl-6} and Figure~\ref{fig-4}) provides the total separation, a, and $\alpha_A$, the perturbation orbit size.
Because at each instant
in the orbits of the two components around the common center of mass

\beq
{\cal M}_A / {\cal M}_B = \alpha_B / \alpha_A
\eeq

we can express the mass fraction

\beq
f = {\cal M}_B/({\cal M}_A+{\cal M}_B) = \alpha_A / (\alpha_A + \alpha_B),
\eeq
where $ \alpha_B = $ a $-\alpha_A.$
From the three epochs of component B detection we derive a mass fraction of 0.3051 $\pm$
0.0031. From the orbit parameters in Table~\ref{tbl-5},
$\alpha_A = 33.2 \pm 0.3$ mas.  For comparison, the
semi-major axis of the photocentric orbit was found to be 29 mas
by \cite{Har71}, and 28 $\pm$ 2 mas by \cite{Her78},
although our new value has an error $\sim$10 times smaller.

Our measured semi-major axis (a= 108.8$\pm$0.7 mas), the period $P= 1.4731 \pm 0.0008$ y, and the
absolute parallax $112.9 \pm 0.3$ mas, provide an orbit semi-major
axis of 0.963 $\pm$ 0.007 AU, and a total system mass ${\cal M}_{tot}
= 0.412 \pm 0.009{\cal M}_{\sun}$ from Kepler's 3d law.  From
the mass fraction we derive
${\cal M}_A = 0.286 \pm 0.006{\cal M}_{\sun}$ and ${\cal M}_B= 0.126
\pm 0.003 {\cal M}_{\sun}$, which are masses with only a 2\%
formal error.

\section{Gl 791.2 A and B on the MLR}

With the component masses determined, we require the component
absolute magnitudes to place these stars on the MLR.  We use the photometry of both components from \cite{Leg92},
$V = 13.06 \pm 0.03$, our measured $\Delta V$ = 3.27 $\pm$ 0.10, and
our parallax, $\pi_{abs} = 112.9 \pm 0.3$ mas to derive
M$_{VA} = 13.37 \pm 0.03$ and M$_{VB} = 16.64\pm 0.10$.  For this very nearby system we
have assumed no absorption ($A_V = 0$).  We collect all derived mass
and absolute magnitude values in Table~\ref{tbl-7}.

\begin{figure}
\epsscale{0.7}
\plotone{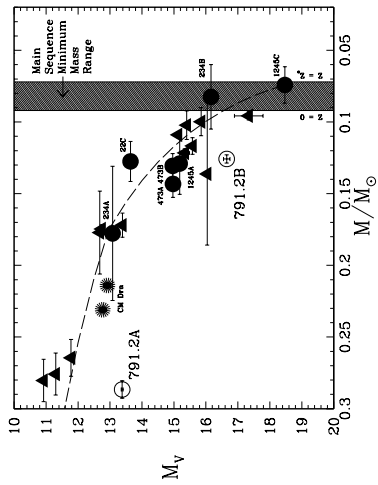}
\caption{ The components of Gl 791.2 are shown on the mass-luminosity diagram.
Round points (labeled) indicate masses determined by this group, while
triangular points rely upon orbits and masses determined by others.
The two components of the eclipsing binary CM Dra are shown with
starred points.  The dashed curve is the empirical mass-luminosity
relation from Henry and McCarthy (1993) down to 0.18 ${\cal M}_{\sun}$ and
from Henry et al (1999) at lower masses.  The shaded region with
borders at 0.092 and 0.072 ${\cal M}_{\sun}$ marks the main-sequence minimim
mass range for objects with zero to solar metallicity.} \label{fig-5}

\end{figure}

Components A and B lie on the MLR as shown in Figure~\ref{fig-5}, with
Gl 791.2 B one of the lowest mass objects on the lower main sequence. The lower main sequence relation is from Henry et al. (1999). The straight-line higher mass section (${\cal M} > 0.2{\cal M}_{\sun}$) is from Henry \& McCarthy (1993).
The distance between the present locus of the 
MLR and our results for Gl 791.2 requires some comment. We first suggest that Gl 791.2 A and B are underluminous. This is possible if they have metallicity higher than solar. Recent M dwarf models (\cite{Brf98}) show that the absolute magnitude of an M 
dwarf star depends not only on mass, but also on age (evolutionary stage) and
very sensitively on metallicity. Higher metallicity stars have fainter M$_V$ than lower metallicity stars. This is the explanation offered by \cite{Del99}, for Gl 866 ABC, also subluminous compared to their MLR. The rapid rotation (V$_{rot} = 32$ km s$^{-1}$) and young disk kinematics of Gl 791.2 (\cite{Del98}) support relatively recent formation and, hence, probably higher than solar metallicity. Second, because the astrometric residuals for Gl 791.2A (Table~\ref{tbl-2} and Figure~\ref{fig-3}) are larger than we typically
obtain, some of the perceived underluminosity might rather be a consequence of excess mass, due to an undetected, short-period low mass companion.

\section{Conclusions}

1. Observations of the low-mass binary Gl 791.2 with {\it HST} FGS 3
over a 1.7 year time span yield a parallax with an external error
better than 1\%.

2. Fringe tracking observations of the primary, component A, provide a
well-determined perturbation orbit. In this case a large $\Delta V$
produces very small and correctable effects on the interferometer response
function, permitting high precision positional measurements.

3. Demonstrating the difficulty of measuring both components of large
$\Delta V$ targets with {\it HST}, fringe scanning observations
succeeded only about 20\% of the time, but provide a mass fraction =
0.3051 $\pm$0.0031 and $\Delta V = 3.27 \pm 0.10$. The component
magnitude difference was only obtained at epochs of largest
separation.

4. We obtain masses precise to 2\%: ${\cal M}_{tot} = 0.412 \pm
0.009{\cal M}_{\sun}$, ${\cal M}_A = 0.286 \pm 0.006{\cal M}_{\sun}$
and ${\cal M}_B= 0.126 \pm 0.003 {\cal M}_{\sun}$.

5. These masses and absolute magnitudes assist in
defining the lower main sequence MLR. The dependence of the lower main sequence MLR on metallicity is an important question for the future.

\acknowledgments

This research has made use of NASA's Astrophysics Data System Abstract
Service and the SIMBAD Stellar Database inquiry and retrieval
system. Web access to the astrometry and photometry in USNO-A2.0 is
gratefully acknowledged. Support for this work was provided by NASA
through grants GTO NAG5- 1603, GO-06036.01-94A and GO-07491.01-97A
from the Space Telescope Science Institute, which is operated by the
Association of Universities for Research in Astronomy, Inc., under
NASA contract NAS5-26555.  Denise Taylor provided crucial scheduling
assistance at the Space Telescope Science Institute. We thank Thierry Forveille
for an early copy of the Gl 866 results. An anonymous referee
made several suggestions that prompted a revisit to these data, resulting in a better paper.

\clearpage

\begin{center}
\begin{deluxetable}{lll}
\tablecaption{ Gl 791.2 = G 24-16 = LHS 3556 = HU Del \label{tbl-1}}
\tablewidth{0in}
\tablehead{\colhead{Parameter} &  \colhead{Value}&
\colhead{Reference}}
\startdata
$V$ & 	   	13.06$\pm$0.03 & 			\cite{Leg92} 	  	\nl
$(B-V)$ &   	1.66 $\pm$ 0.05 & 			\cite{Leg92} 		\nl
Sp.Type & 	M4.5V &    				\cite{Kir91}		\nl
$R_A$ &  	$0.18R_{\sun}$& 			\cite{Bur93}		\nl
$V_A(rot)$ & 	$32 \pm 2$ km s$^{-1}$ & 		\cite{Del98}		\nl
\enddata
\end{deluxetable}
\end{center}

\clearpage

\begin{center}
\begin{deluxetable}{lcccccc}
\tablecaption{POS Observations of Gl 791.2A and Perturbation Orbit Residuals \label{tbl-2}}
\tablewidth{0in}
\tablehead{\colhead{Obs. Set} &  \colhead{ MJD }&  \colhead{ $\rho_A$ (mas) }&  \colhead{ $\theta$ (\arcdeg) }&  \colhead{ $\Delta\rho_A$ (mas)}&  \colhead{$\Delta\theta$ (\arcdeg) }&  \colhead{ $\rho_A\Delta\theta$ (mas) }}
\scriptsize
\startdata
1&49950.1779&16.0&95.1&0.3&5.6&1.6\\
1&49950.1952&15.9&95.1&0.4&5.0&1.4\\
2&49970.9509&16.1&139.8&0.4&-2.3&-0.7\\
2&49970.9682&16.1&139.8&0.4&-5.1&-1.4\\
3&49997.8257&21.4&183.6&-0.9&-8.0&-3.0\\
3&49997.8430&20.9&183.6&-0.4&-7.1&-2.6\\
4&50025.5061&26.8&208.6&0.0&1.5&0.7\\
4&50025.5234&27.3&208.6&-0.6&0.0&0.0\\
5&50050.6372&32.4&222.5&0.0&5.3&3.0\\
5&50050.6545&32.7&222.5&-0.3&5.0&2.8\\
6&50167.3154&48.0&256.1&-0.5&2.1&1.8\\
6&50167.3327&49.8&256.1&-2.3&3.0&2.6\\
7&50194.5250&48.8&261.8&-0.6&1.9&1.6\\
7&50194.5423&47.7&261.8&0.5&0.9&0.8\\
8&50236.9474&47.0&270.7&0.5&-1.0&-0.8\\
8&50236.9647&47.3&270.7&0.2&-0.2&-0.2\\
9&50255.2418&49.5&274.7&-3.0&-1.6&-1.4\\
9&50255.2591&49.7&274.7&-3.3&-1.1&-1.0\\
10&50338.7445&38.2&297.1&-1.4&-1.7&-1.2\\
10&50338.7618&37.5&297.1&-0.7&2.3&1.5\\
10&50338.7684&38.6&297.1&-1.8&-1.3&-0.9\\
11&50372.9877&33.6&311.3&-2.8&3.4&2.0\\
11&50373.0050&33.4&311.3&-2.6&3.6&2.1\\
11&50373.0117&32.7&311.3&-2.0&5.2&3.0\\
12&50398.5885&25.9&326.5&-0.1&8.4&3.8\\
12&50398.6058&26.3&326.5&-0.5&8.0&3.7\\
12&50398.6125&27.0&326.5&-1.1&7.9&3.7\\
13&50424.6554&21.2&349.5&0.1&5.6&2.1\\
13&50424.6727&22.1&349.5&-0.8&5.9&2.3\\
13&50424.6793&23.0&349.5&-1.7&3.4&1.4\\
14&50569.0266&28.3&212.1&-0.5&-2.2&-1.1\\
14&50569.0439&29.1&212.1&-1.3&-3.3&-1.7\\
14&50569.0505&29.7&212.1&-1.9&-3.8&-2.0\\
& N = 33& $<|res|>$& =  &1.0&3.7&1.8\\
\enddata
\end{deluxetable}
\end{center}

\clearpage
\begin{center}
\begin{deluxetable}{lcccccccc}
\tablecaption{TRANS Observations of Gl 791.2B and Residuals \label{tbl-3}}
\tablewidth{0in}
\tablehead{\colhead{Obs. Set} &  \colhead{ MJD }&  \colhead{ $\rho_A$ (mas) }&  \colhead{ $\theta$ (\arcdeg) }&  \colhead{ $\Delta\rho_A$ (mas)}&  \colhead{$\Delta\theta$ (\arcdeg) }&  \colhead{ $\rho_A\Delta\theta$ (mas) }&\colhead{ $\Delta m_X^c$}&\colhead{ $\Delta m_Y$}}
\scriptsize
\startdata
8&50235.94744&169.2&89.2&7.4&0.0&0.0&3.40&3.12\\
9&50254.24177&153.3&84.6&-5.2&-0.7&-1.8&3.54&3.43\\
10&50337.74448&126.0&62.9&-0.3&0.3&0.2&3.14&3.30\\
 & N = 3& &$<|res|>=$&4.3&0.3&0.7&3.36&3.28\\ 
\enddata
\end{deluxetable}
\end{center}

\clearpage

\begin{center}
\begin{deluxetable}{lcccccc}
\tablewidth{0in}
\tablecaption{Gl 791.2 Reference Frame Relative Positions \label{tbl-4}}
\tablehead{  \colhead{ID}&\colhead{V}& \colhead{\bv} & \colhead{$\xi$}&  
&\colhead{$\eta$}&   
\\ & & &   (arcsec)& & 
(arcsec) &}

\startdata
ref-2&12.1&1.9&109\farcs8430&$\pm$0\farcs0009&29\farcs0122&$\pm$0\farcs0011\\
ref-3&14.8&1.5&78.4308&0.0010&5.4658&0.0011\\
ref-4*&14.8&1.6&0.0000&0.0017&0.0000&0.0023\\
\enddata
\tablenotetext{*} {RA, Dec = 307\fdg43704, 9\fdg68324 (J2000, epoch 1995.843)}
\end{deluxetable}
\end{center}

\clearpage

\begin{center}
\begin{deluxetable}{lll}
\tablecaption{Gl 791.2 Parallax and Proper Motion \label{tbl-5}}
\tablewidth{0in}
\tablehead{\colhead{Parameter} &  \colhead{ Value }}
\startdata

{\it HST} study duration &	1.7 yr				\nl
number of observation sets &   	14 			\nl
ref. stars $ <V> $ &  		$13.9 \pm 1.6$  		\nl
ref. stars $ <B-V> $ &  	$1.7 \pm 0.2$ 			\nl
&								\nl
{\it HST} Relative Parallax &	111.9 $\pm$ 0.2 mas		\nl
correction to absolute &	1.0 $\pm$ 0.2 mas		\nl
{\it HST} Absolute Parallax & 	112.9 $\pm$ 0.3 mas		\nl
{\it HST} Proper Motion &	678.8 $\pm$ 0.4	mas y$^{-1}$    \nl
\indent in pos. angle & 		79\fdg62 $\pm$ 0\fdg03		\nl
&								\nl
Yale Parallax Catalog (1995) & 	113.8 $\pm$ 1.9 mas		\nl
Proper Motion (\cite{Har80}) &	678.6 mas y$^{-1}$ 		\nl
\indent in pos. angle & 		79\fdg0 			\nl
\enddata
\end{deluxetable}
\end{center}

\clearpage

\begin{center}
\begin{deluxetable}{cl}
\tablecaption{ Gl 791.2AB Orbit\label{tbl-6}}
\tablewidth{0in}
\tablehead{\colhead{Parameter} &  \colhead{Value}}
\startdata
a&	108.8$\pm$0.7 mas	\\
$\alpha_A$&		33.2$\pm$0.2 mas			\\
$\alpha_B$&		75.6$\pm$0.7 mas			\\
f&0.3051 $\pm$0.0031\\
a& 			$0.963\pm 0.007$ AU	\\
P&			1.4731 $\pm$0.0008 y	\\
T&			1998.5668$\pm$0.0013	\\
e&			0.519$\pm$0.003	\\
i&			145\fdg7$\pm$0\fdg7\\
$\Omega^{\prime}$&	104\fdg5$\pm$0\fdg2	\\
$\Omega$&	-75\fdg5$\pm$0\fdg2		\\
$\omega^{\prime}$&	12\fdg8$\pm$0\fdg3\\ 
\enddata
\end{deluxetable}
\end{center}

\clearpage

\begin{center}
\begin{deluxetable}{cc}
\tablecaption{Gl 791.2: Component Masses and M$_V$ \label{tbl-7}}
\tablewidth{0in}
\tablehead{\colhead{Parameter} &  \colhead{Value}}
\startdata
${\cal M}_{tot} $&$ 0.412 \pm 0.009~{\cal M}_{\sun}$\\
${\cal M}_A $&$ 0.286 \pm 0.006~{\cal M}_{\sun}$\\
${\cal M}_B$&$ 0.126 \pm 0.003~{\cal M}_{\sun}$\\
$M_A $&$ 13.37 \pm 0.03 $\\
$M_B $&$16.64 \pm 0.10$\
\enddata
\end{deluxetable}
\end{center}

\clearpage

%

\clearpage

\clearpage

\clearpage

\clearpage

\clearpage



\end{document}